\documentclass[10pt,a4paper,journal,comsoc]{IEEEtran}
\IEEEoverridecommandlockouts
\usepackage[a4paper,
            left=0.7in,right=.75in,top=0.75in,bottom=0.7in]{geometry}
\usepackage{graphicx}
\usepackage{color}
\usepackage{cite}
\usepackage{amsmath}
\usepackage[caption=false]{subfig}
\usepackage{mathtools}
\usepackage{amssymb}
\usepackage{amsfonts}

\usepackage{algorithmic}
\usepackage{algorithm}

\usepackage{amsmath}

\usepackage{url}
\newcommand{\subparagraph}{}
\usepackage{amssymb}
\usepackage{amsfonts}
\usepackage{lineno}
\usepackage{comment}

\usepackage{siunitx}

\usepackage{paralist} % Beware of enu­mer­ated and item­ized lists. Instead, replace them with com­pact lists.

\newtheorem{remark}{Remark}

\newcounter{tempEquationCounter}
\newcounter{thisEquationNumber}

\begin{document}
\vspace{-5.0cm}
\title{ \LARGE Ultra-reliable communication in 5G mmWave networks: \\A risk-sensitive approach}
\author{Trung Kien Vu,~\IEEEmembership{Student Member,~IEEE,}
        Mehdi Bennis,~\IEEEmembership{Senior Member,~IEEE,} \\
        M\' erouane Debbah,~\IEEEmembership{Fellow,~IEEE,}
        Matti Latva-aho,~\IEEEmembership{Senior Member,~IEEE,}\\
        and~Choong Seon Hong,~\IEEEmembership{Senior Member,~IEEE}
\vspace{-1cm}
\thanks{Manuscript received Dec 29, 2017; revised Jan 20, 2018; accepted Jan 31, 2018. Date of publication Feb 06, 2018; This work was supported in part by Tekes, Nokia, Huawei, MediaTek, Keysight, Bittium, Kyynel, in part by the Academy of Finland via the grant 307492 and the CARMA grants 294128 and 289611. The Nokia Foundation and the Tekniikan edistanmissäätiö are also acknowledged. The associate editor coordinating the review of this letter and approving it for publication was A. S. Cacciapuoti. (Corresponding author: Trung Kien Vu.)}
%CARMA project, and

\thanks{T. K. Vu and M. Latva-aho are with the Centre for Wireless Communications, University of Oulu, Oulu 90014, Finland (e-mail: trungkien.vu@oulu.fi; matti.latva-aho@oulu.fi).}

\thanks{M. Bennis is with the Centre for Wireless Communications, University of Oulu, 90014 Oulu, Finland, and also with the Department of Computer Science and Engineering, Kyung Hee University, Yongin 446-701, South Korea (e-mail: mehdi.bennis@oulu.fi).}

\thanks{M. Debbah is with the Large Networks and System Group (LANEAS), CentraleSup\'elec, Universit\'e Paris-Saclay, 91192 Gif-sur-Yvette, France and also with the Mathematical and Algorithmic Sciences Laboratory, Huawei France R\&D, 92100 Paris, France  (e-mail: merouane.debbah@huawei.com).}

\thanks{C. S. Hong is with the Department of Computer Science and Engineering, Kyung Hee University, Yongin 446-701, South Korea (email: cshong@khu.ac.kr).}

\thanks{Citation information: 10.1109/LCOMM.2018.2802902, IEEE Communications Letters, 21 (9), 2041--2044.}

}\vspace{-5.0cm}\maketitle
\begin{abstract}
	This letter investigates the problem of providing gigabit wireless access with reliable communication in $5$G millimeter-Wave (mmWave)
	massive multiple-input multiple-output (MIMO) networks. In contrast to the classical network design based on average metrics, a distributed \textit{risk-sensitive} reinforcement learning-based framework is proposed to jointly optimize the beamwidth and transmit power, while taking the sensitivity of mmWave links  into account. Numerical results show that the proposed algorithm achieves more than $9$ Gbps of user throughput with a guaranteed probability of $90\%$, whereas the baselines guarantee less than $7.5$ Gbps. More importantly, there exists a rate-reliability-network density tradeoff, in which as the user density increases from $16$ to $96$ per $\text{km}^{2}$, the fraction of users that achieve $4$ Gbps are reduced by $11.61\%$ and $39.11\%$ in the proposed and the baseline models, respectively.
\end{abstract}
\IEEEpeerreviewmaketitle{}
    
    \begin{IEEEkeywords}
    	URLLC, URC, reliable communication, mmWave communications, risk-sensitive learning,	reinforcement learning.
    \end{IEEEkeywords}

\section{Introduction}
\label{introduction}
   To enable gigabit wireless access with reliable communication, a number of candidate solutions are currently investigated for $5$G: $1$) higher frequency spectrum, e.g., millimeter wave (mmWave); $2$) advanced spectral-efficient techniques, e.g., massive multiple-input multiple-output (MIMO); and	$3$) ultra-dense small cells \cite{5GWhat}. This work explores the above techniques to enhance the wireless access~\cite{5GWhat, 2015SmallCell, Vu2017}. Massive MIMO  yields remarkable properties such as high signal-to-interference-plus-noise ratio due to large antenna gains, and extreme spatial multiplexing gain \cite{Vu2017,2016secure}. Specially, mmWave frequency bands offer huge bandwidth \cite{2013millimeter}, while it allows for packing a massive antennas for highly directional beamforming~\cite{2013millimeter}. A unique peculiarity of mmWave is that mmWave links are very sensitive to blockage, which gives rise to unstable connectivity and unreliable communication \cite{2013millimeter}. To overcome such challenge,  this letter applies  principles of risk-sensitive reinforcement learning (RSL) and exploits the multiple antennas diversity and higher bandwidth to optimize transmission to achieve gigabit data rates, while considering the sensitivity of mmWave links to provide ultra-reliable communication (URC). The prime motivation behind using RSL stems from the fact that the \textit{risk-sensitive} utility function to be optimized is a function of not only the average but also the variance \cite{2002risk}, and thus it captures the tail of rate distribution to enable URC. While the proposed algorithm is fully distributed, which does not require  full network observation, and thus the cost of channel estimation and signaling synchronization is reduced. Via  numerical experiments, we showcase the inherently key trade-offs between ($i$) reliability/data rates and network density, and ($ii$) availability and network density.

   %Massive MIMO  yields remarkable properties such as high signal-to-interference-plus-noise ratio due to large antenna gains, and extreme spatial multiplexing gain \cite{Vu2017,2014mmW}, \cite{2016secure}

\textbf{Related work}: In \cite{2017ultra_petar, 2018ultra}  authors provided
	the principles of ultra-reliable and low latency communication (URLLC) and described some techniques to support URLLC. Recently, the problem of low latency communication \cite{2017low} and URLLC \cite{vu2017ultra, 2018Path} for $5$G mmWave network was studied to evaluate the  performance under the impact of traffic dispersion and network densification. Moreover, a reinforcement learning (RL) approach to power control and rate adaptation was studied in \cite{2017RL}. All these works  focus on maximizing the time average of network throughput or minimizing the mean delay without providing any guarantees for higher order moments (e.g., variance, skewness, kurtosis, etc.). This work departs from the classical average-based system design and instead takes higher order moments in the utility function into account to formulate a RSL framework through which every small cell optimizes its transmission while mitigating signal fluctuations.

\section{System Model}
\label{SystemModel}
Let us consider a mmWave downlink (DL) transmission of a small cell network
consisting of a set ${\cal B}$ of $B$ small cells (SCs), and a set ${\cal K}$ of $K$ user equipments (UEs) equipped with $N_{k}$ antennas.
We assume that each SC is equipped with a large number of $N_{b}$ antennas  to exploit massive MIMO gain and adopt a hybrid beamforming architecture \cite{2017hybrid}, and  assume that $N_{b}\gg N_{k}\geq1$ . Without loss of generality, one UE per one SC is considered\footnote{For the multiple UEs case,  addition channel estimation and user scheduling need to be considered, one example was studied in \cite{Vu2017}.}. The data traffic is generated from SC to UE via mmWave communication. A co-channel time-division duplexing protocol is considered, in which the DL channel can be obtained via the uplink training phase.

Each SC adopts the hybrid beamforming architecture, which enjoys both analog and digital beamforming techniques \cite{2017hybrid}. Let $g_{bk}^{(tx)}$ and $g_{bk}^{(rx)}$ denote the analog transmitter and receiver beamforming gains at the SC $b$ and UE $k$, respectively.
In addition, we use $\omega_{bk}^{(tx)}$ and $\omega_{bk}^{(rx)}$ to represent the angles deviating from the strongest path between the SC $b$ and UE $k$. Also, let $\theta_{bk}^{(tx)}$ and $\theta_{bk}^{(rx)}$ denote the beamwidth at the SC and UE, respectively. We denote $\boldsymbol{\theta}$ as a vector of the transmitter beamwidth of all SCs. We adopt the widely used antenna radiation pattern model \cite{2017hybrid} to determine the analog beamforming gain as
\begin{align}
g_{bk}\left(\omega_{bk},\theta_{bk}\right) & =\begin{cases}
\frac{2\pi-\left(2\pi-\theta_{bk}\right)\eta}{\theta_{bk}}, & \text{if}\:|\omega_{bk}|\leq\frac{\theta_{bk}}{2},\\
\eta, & \text{otherwise,}
\end{cases}\label{eq:beamgain}
\end{align}
where $0<\eta\ll1$ is the side lobe gain.

 Let ${\bf H}_{bk}\in\mathbb{C}^{N_{b}\times N_{k}}$ denote the channel propagation matrix (channel state) from SC $b$ to UE $k$. We assume a time-varying channel state described by a Markov chain and there are $T \in \mathcal{Z}^{+}$ states, i.e., for each ${\bf H}_{bk}(t), t = \{1, \ldots, T\}$.  Considering the imperfect channel state information (CSI), the estimated channel state between the SC $b$  and  UE $k$  is modelled as \cite{vu2017ultra}
\begin{equation}
\hat{\bf H}_{bk}=  \sqrt{N_{b}\times N_{k}} \boldsymbol{\Theta}_{bk}^{1/2} \Big(  \sqrt{1 - \tau_k^2} {\bf W}_{bk} + \tau_k \hat{\bf W}_{bk} \Big), \notag
\end{equation}
where $\boldsymbol{\Theta}_{bk}\in\mathbb{C}^{N_{b}\times N_{b}}$ is the spatial channel correlation matrix with a low rank that accounts for the mmWave channel path loss and shadow fading \cite{2014mmWave_Blockage,Vu2016}. Moreover, the spatial channel model is clustered, which belongs to a finite set with a finite size \cite{Liu2014}. Here, ${\bf W}_{bk}\in\mathcal{C}^{N_{b}\times N_{k}}$ is the small-scale fading channel matrix, modelled as a random matrix with a zero mean and a variance of $\frac{1}{N_{b}\times N_{k}}$. Here $\tau_k \in [0, 1]$ reflects the estimation accuracy for UE $k$, if $\tau_k = 0$, and we assume perfect channel state information. $\hat {\bf W}_{bk} \in \mathbb{C} ^{N_{b} \times N_{k}}$ is the estimated noise vector, also modeled as a random matrix with a zero mean and a variance of $\frac {1} {N_{b} \times N_{k}}$. We denote ${\bf H}= \{ {\bf H}_{bk} |  \forall b  \in {\mathcal B} , \forall k  \in {\mathcal K} \} $ as the network state.% containing all CSI.

By applying a linear precoding scheme ${\bf V}_{bk}(\hat{\bf H}_{bk})$ \cite{2017hybrid}, i.e, ${\bf V}_{bk}(\hat{\bf H}_{bk})=\hat{\bf H}_{bk}$ for the conjugate precoding, the achievable rate\footnote{Note that we omit the beam search/track time, since it can be done in a short time compared to transmission time \cite{2017tracking}. We assume that each BS sends a single stream to its users via the main beams.} of UE $k$ from SC $b$ can be calculated as
\begin{align*}
r_{b}\left(t\right) & =\text{w}\log \Big (1+\frac{p_{b}g_{bk}^{(tx)}g_{bk}^{(rx)}|{\bf H}_{bk}^{\dagger}{\bf V}_{bk}|^{2}}{\sum_{b^{\prime}\neq b} p_{b^{\prime}} g_{b^{\prime}k}^{(tx)}g_{b^{\prime}k}^{(rx)}|{\bf H}_{b^{\prime}k}^{\dagger}{\bf V}_{b^{\prime}k}|^{2} + \sigma^2_{bk} } \Big),
\end{align*}
where $p_{b}$ and $p_{b^{\prime}}$ are the transmit powers of SC $b$ and SC $b^{\prime}$, respectively. In addition, $\text{w}$ denotes the system bandwidth of the mmWave frequency band. The thermal noise of user $k$ served by SC $b$ is $\eta_{bk} \sim \mathcal{CN}(0,\sigma^2_{bk})$ . Here, we denote $P_{b}^{{\rm max}}$ as the maximum transmit power of SC $b$ and $\mathbf{p}=(p_{b}|\forall b\in{\mathcal B},\:0\leq p_{b}\leq P_{b}^{\max})$ as the transmit power vector.
\section{Problem Formulation}
\label{Pro-Form}
We model a decentralized optimization problem and harness tools from RSL to solve, whereby SCs autonomously respond to the network states based on the historical data. Let us consider a joint optimization of transmitter beamwidth\footnote{As studied in \cite{2017hybrid}, for $\eta\leq\frac{1}{3}$, the problem of selecting beamwidth for the transmitter and receiver can be done by adjusting the transmitter beamwidth with a fixed receiver beamwidth.} $\boldsymbol{\theta}$ and transmit power allocation ${\bf p}$. We denote ${\bf z} \left( t \right) = \left( \boldsymbol{ \theta} \left( t \right) , {\bf p} \left (t\right) \right)$, which takes values in ${\cal Z} = \left\{ {\bf z}_{1}, \cdots, {\bf z}_{B}\right\} $, where ${\bf z}_{b}=\left(\theta_{b},\:p_{b}\right)$. Assume that each SC $b$ selects its beamwidth and transmit power drawn from  a given probability distribution $\boldsymbol {\pi}_{b} = \big(\pi_{b}^{1}, \cdots, \pi_{b}^{m}, \cdots, \pi_{b}^{Z_{b}} \big)$ in which $Z_{b}$ is the cardinality of the set of all combinations $\left(\theta_{b},\:p_{b}\right)$, i.e.,  $\sum _{m=1}^{Z_{b}} \pi_{b}^{m} = 1$. For each $m=\left\{ 1,\cdots,Z_{b}\right\} $ and ${\bf z}_{b}^{m} = (\theta_{b}^{m}, \:p_{b}^{m})$ the mixed-strategy probability is defined as
\begin{align}
\pi_{b}^{m}(t)=\text{Pr}({\bf z}_{b}(t) = {\bf z}_{b}^{m} | {\bf z}_{b}(0:t-1) , {\boldsymbol \pi}_{b}(0:t-1) ).
\label{policy}
\end{align}
We denote $\boldsymbol{\pi} = \{\boldsymbol{\pi}_{1}, \cdots, \boldsymbol{\pi}_{b}, \cdots, \boldsymbol{\pi}_{B}\} \in \Pi$, in which $\Pi$ is the set of all possible probability mass functions (PMF). Let ${\bf r}=({\bf r}_{1},\cdots,{\bf r}_{B})$ denote the instantaneous rates, in which ${\bf r}_{b}=(r_{b}(0),\cdots,r_{b}(T))$.
Let $\mathcal{R}$ denote the rate region, which is defined as the convex hull of the rates \cite{2004convex}, i.e., ${\bf r}\in\mathcal{R}$. Inspired by the RSL \cite{2002risk}, we consider the following utility function, given by
\begin{equation}
\bar{u}_{b}=\frac{1}{\mu_{b}}\log\mathbb{E}_{{\bf H},\boldsymbol{\pi}}\left[\exp(\mu_{b}\sum_{t=0}^{T}r_{b}(t))\right],
\label{reward}
\end{equation}
where the parameter $\mu_{b}<0$ denotes the desired risk-sensitivity, which will \textit{penalize the variability} \cite{2002risk} and the operator $\mathbb{E}$ denotes the expectation operation.
\begin{remark}\label{remark1}
	The Taylor expansion of the utility function given in \eqref{reward} yields
	\[
	\bar{u}_{b}\triangleq\mathbb{E}_{{\bf H},\boldsymbol{\pi}}\left[\sum_{t=0}^{T}r_{b}(t)\right]+\frac{\mu_{b}}{2}\text{Var}_{{\bf H},\boldsymbol{\pi}}\left[\sum_{t=0}^{T}r_{b}(t)\right]+\mathcal{O}\left(\mu_{b}^{2}\right).
	\]
\end{remark}
\textit{Remark 1} basically shows that the utility function \eqref{reward} considers both mean and variance terms ($\text{Var}$) of the mmWave links. We formulate the following distributed optimization problem for every SC as:
\begin{subequations}\label{OP1}
	\begin{eqnarray}
	\max_{\boldsymbol{\pi_{b}}} &  & \frac{1}{\mu_{b}}\log\mathbb{E}_{{\bf H},\boldsymbol{\pi_{b}}}\left[\exp(\mu_{b}\sum_{t=0}^{T}r_{b}(t))\right]\\
	\text{subject to} &  & {\bf r}_{b}\in\mathcal{R},\:\boldsymbol{\pi}_{b}\in\Pi,\:p_{b}\leq P_{b}^{\max}.
	\end{eqnarray}
	
\end{subequations}
It is challenging to solve \eqref{OP1} if each SC does not have full network observation. This work does not assume an explicit knowledge of the state transition probabilities. Here, we leverage principles of RL to optimize the transmit beam in a totally decentralized manner \cite{2002risk, 2017RL, 2013learning}.
    \vspace{-0.5em}
\section{Proposed Algorithm}
\label{Design}
This section introduces reinforcement learning tool used to address the pre-defined problem \eqref{OP1}. A learning prodedure is then proposed to refine and solve \eqref{OP1}. Finally, the covergence conditions for the learning rates are established.
\subsection{Introduction to Reinforcement Learning}
\label{IRL}
Reinforcement learning is an area of machine learning in which agents perform actions to interact with the environment so as to maximize the cumulative reward \cite{2011learning}. By evaluating feedback from theirs own actions and experiences, the agents determine a sequence of best actions which maximize the long-term reward.

\begin{figure}[t]
	\centering
	\includegraphics[width=1\columnwidth]{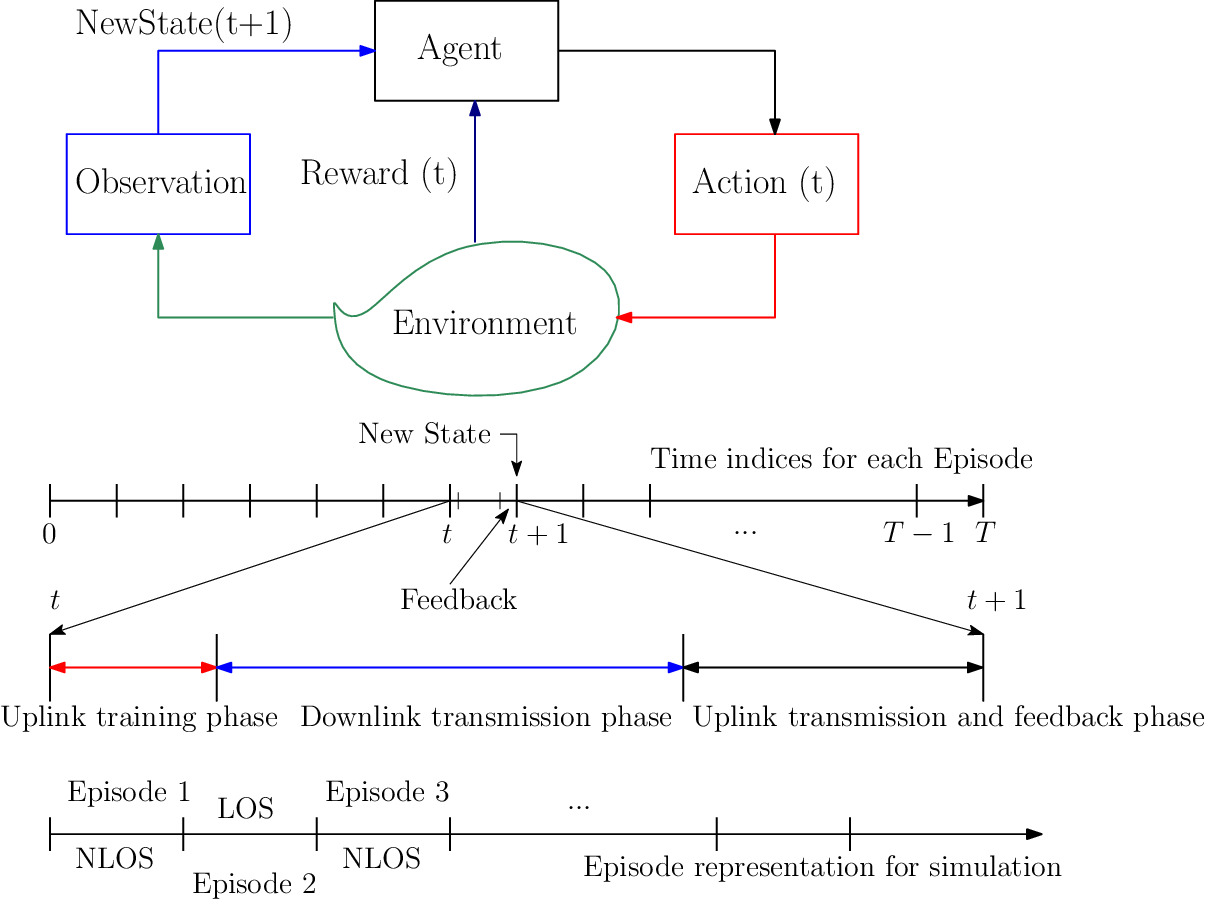}
	\vspace{-1em}
	\caption{A reinforcement learning model.}
	\vspace{-1em}
	\label{RL_Model_manus}
\end{figure}

Basically, reinforcement learning is concerned with decision making to enable the adaptation and self-organization, and the agents spend time discovering actions to find the best strategies, then exploit them in the long run. At each time slot $t$, each agent selects an action from a possible action set, the agent observes the environment and experiences the reward  as shown in Fig. \ref{RL_Model_manus}. In the next time slot $t+1$, the agent evaluates the decision, which is made from the previous time slot and the agent selects the action based on the distribution of the action-reward. Here, the concept of regret strategy is employed, defined as the difference between the average utility when choosing the same actions in previous times, and its average utility obtained by constantly selecting different actions. The premise is that regret should be minimized over time so as to choose the best sequence of actions.

The important elements of reinforcement learning include agents,  actions,  reward function, policy and  environment, which are  briefly described as follows:
\begin{itemize}
	\item \textbf{Agents} can be network operators, base stations, or users, who want to maximize their cumulative reward functions.
	
	\item \textbf{Actions} are defined as a set of things that agents do to solve their concerns with the environments. In the context of resource allocation, actions could consist of user association, power assignment, or beamwidth selection.
	
	\item \textbf{Reward} function is defined as the cumulative return for the agent after applying selected actions to the environment. Network utility function and power consumption are common metrics used to measure the reward.
	
	\item \textbf{Policy} refers to strategies that  the agents play to determine next action based on the distribution of actions-rewards. It is a mapping between  action and  state. Here, a state is the current condition of the environment such as the channel state, or network queuing state.
	
	\item \textbf{The environment} contains the network system, where the agents play their actions to maximize the reward. At the beginning of each time slot, the agents observe the reward, which reflects the noise and interference in the environment.
	
\end{itemize}

\subsection{Proposed Algorithm}
\label{IRL}
We leverage the reinforcement learning tool to solve the predefined problem. In particular, each SC acts as an agent which selects an action to maximize a long-term reward based on user feedback and probability distribution for each action. The action is defined as the selection of ${\bf z}_{b}$, while the long-term utility in \eqref{OP1} is the reward, and the environment here contains the network state. To this end, we build the probability distribution for every action and provide a RL procedure to solve \eqref{OP1}.

We denote $u_{b}^{m}=u_{b}^{m}\left({\bf z}_{b}^{m},{\bf z}_{-b}\right)$ as a utility function of SC $b$ when selecting ${\bf z}_{b}^{m}$. Here, ${\bf z}_{-b}$ denotes the composite variable of other agents' actions excluding SC $b$. From \eqref{reward}, the utility $u_{b}\left(t\right)$ of SC $b$ at time slot $t$, i.e., $\bar{u}_{b} = \sum_{t=0}^{T}u_{b} \left (t \right )$, is rewritten as
\begin{equation}
u_{b}\left(t\right)=\frac{1}{\mu_{b}}\log\left(\sum_{m=1}^{Z_{b}}\pi_{b}^{m}\exp\left(\mu_{b}r_{b}^{m}\left({\bf z}_{b}^{m}\left(t\right),{\bf z}_{-b}\right)\right)\right),\label{eq:Final_Utility}
\end{equation}
where $r_{b}^{m}({\bf z}_{b}^{m}\left(t\right),{\bf z}_{-b})$ is the instantaneous rate of SC $b$ when choosing ${\bf z}_{b}^{m}\left(t\right)=(\theta_{b}^{m}\left(t\right),\:p_{b}^{m}\left(t\right))$
with probability $\pi_{b}^{m}\left(t\right)$.
\begin{remark}\label{remark2}
For a small $\mu_{b},$ \eqref{reward} is approximated via the Taylor approximation\footnote{For a small $x > 0$, the Taylor approximation of $\log\left(x\right)$ is $x-1$.} of $r_{b}$ around $\mu_{b}\longrightarrow0$ as
\end{remark}
\begin{eqnarray}
\bar{u}_{b} & = & \frac{1}{\mu_{b}}\mathbb{E}\left[\sum_{t=0}^{T}\left(\exp(\mu_{b}r_{b}(t))-1\right)\right],\label{eq:Approximated_Utility}\\
& = & \frac{1}{\left(T+1\right)}\sum_{t=0}^{T}\frac{\exp(\mu_{b}r_{b}(t))-1}{\mu_{b}},\label{eq:FU}
\end{eqnarray}
where \eqref{eq:FU} is obtained by expanding the time average of
\eqref{eq:Approximated_Utility}. Each SC determines $(\theta_{b}^{m},\:p_{b}^{m})$
from ${\cal Z}_{b}$ based on the probability distribution from the
previous stage $t-1$, i.e.,

\begin{align}
{\bf \boldsymbol{\pi}}_{b}\left(t-1\right) & =\left(\pi_{b}^{1}\left(t-1\right),\cdots,\pi_{b}^{Z_{b}}\left(t-1\right)\right).
\label{eq:PreviousProbability}
\end{align}
We introduce the Boltzmann-Gibbs distribution to capture the exploitation and exploration, $\boldsymbol{\beta}_{b}\left({\bf u}_{b}(t)\right)$, given by
\begin{equation}
\begin{alignedat}{1}\boldsymbol{\beta}_{b}^{m}\left({\bf u}_{b}(t)\right)\:=\: & \underset{\boldsymbol{\pi}_{b}\in\Pi}{\mbox{argmax}}\sum_{m\in{\bf z}_{b}}\left[\pi_{b}^{m}u_{b}^{m}\left(t\right)\right.\\
& \qquad\quad\left.-\kappa_{b}\pi_{b}^{m}\ln(\pi_{b}^{m})\right],
\end{alignedat}
\label{eq:GibbsDistribution}
\end{equation}
where ${\bf u}_{b}(t)=\left(u_{b}^{1}\left(t\right),\cdots,u_{b}^{Z_{b}}\left(t\right)\right)$
is the utility vector of SC $b$ for ${\bf z}_{b}\in\mathcal{Z}_{b}$, and the trade-off factor $\kappa_{b}$ is used to balance between exploration and exploitation. If $\kappa_{b}$ is small, the SC selects ${\bf z}_{b}$ with highest payoff. For $\kappa_{b}\rightarrow\infty$ all decisions have equal chance.

For a given ${\bf {\bf u}}_{b}(t)$ and $\kappa_{b}$, we solve  \eqref{eq:GibbsDistribution} to find the probability distribution, by adopting the notion of logit equilibrium \cite{2013learning}, we have
\begin{equation}
\beta_{b}^{m}({\bf {\bf u}}_{b}(t))=\frac{\exp\left(\frac{1}{\kappa_{b}}\left[u_{b}^{m}\right]^{+}\right)}{\sum\limits _{m'\in{\cal Z}_{b}}\exp\left(\frac{1}{\kappa_{b}}\left[u_{b}^{m'}\right]^{+}\right)},
\label{eq:GibbsSolution}
\end{equation}
where $[x]^{+}\equiv\max[x,0]$. Finally, we propose two coupled RL processes that run in parallel and allow SCs to decide their
optimal strategies at each time instant $t$ as follows \cite{2013learning}.

\textbf{\textit{\textcolor{black}{Risk-Sensitive Learning procedure}}}:
We denote $\hat{u}_{b}(t)$ as the estimate utility of SC $b$, in which the estimate utility and probability mass function are updated for each action $m\in Z_{b}$ as follows: 
\[
\begin{cases}
\hat{u}_{b}^{m}\left(t\right)=\hat{u}_{b}^{m}\left(t-1\right)+\\
\quad\quad\quad\quad\zeta_{b}(t)\mathbb{I}_{\{{\bf z}_{b}(t)={\bf z}_{b}^{m}\}}\times\left(u_{b}(t-1)-\hat{u}_{b}^{m}\left(t-1\right)\right),\\
\pi_{b}^{m}\left(t\right)=\pi_{b}^{m}\left(t-1\right)+\iota_{b}(t)\left(\beta_{b}^{m}({\bf u}_{b}(t))-\pi_{b}^{m}\left(t-1\right)\right),
\end{cases}
\]
where $\zeta_{b}(t)$ and $\iota_{b}(t)$ are the learning rates which
satisfy the following conditions (due to space limits please see \cite{2013learning} for convergence proof):
\[
\begin{cases}
\lim_{T\rightarrow\infty}\sum_{t=0}^{T}\zeta_{b}(t) & =+\infty,\lim_{T\rightarrow\infty}\sum_{t=0}^{T}\iota_{b}(t)=+\infty.\\
\lim_{T\rightarrow\infty}\sum_{t=0}^{T}\zeta_{b}^{2}(t) & <+\infty,\lim_{T\rightarrow\infty}\sum_{t=0}^{t}\iota_{b}^{2}(t)<+\infty.\\
\lim_{t\rightarrow\infty}\frac{\iota_{b}(t)}{\zeta_{b}(t)} & =0.
\end{cases}
\]
Finally, each SC determines ${\bf z}_{b}^{m}$
as per \eqref{eq:PreviousProbability}.
\section{Numerical Results}
\label{Evaluation-1}
\begin{figure}[t]
	\includegraphics[width=1\columnwidth]{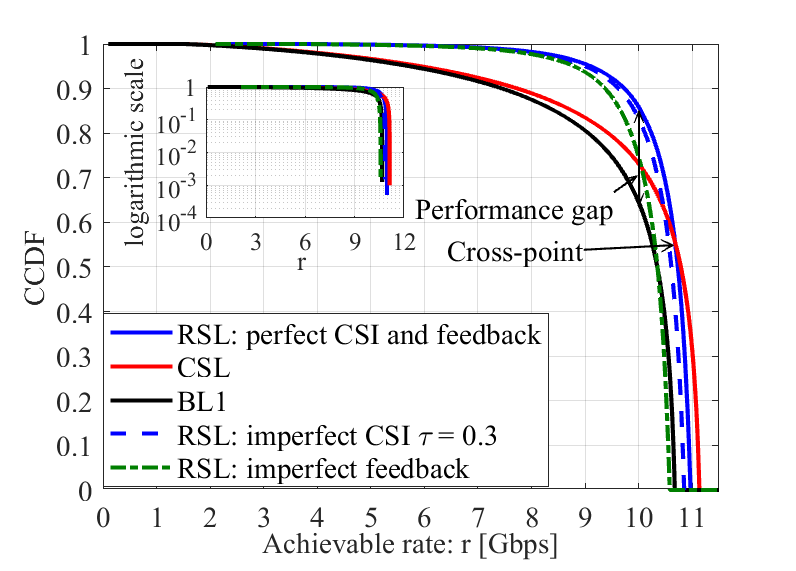}
	\caption{Tail distribution of the achievable rate, $B=24$.}
    \vspace{-1em}
	\label{CCDF}
\end{figure}

A dense SCs are randomly deployed in a $0.5\times0.5$ $\text{km}^{2}$ area and we assume one UE per each
SC and a fixed user association. We assume that each SC adjusts its beamwidth with a step of $0.02$ radian from the range $[\theta^{\text{min}},\:\theta^{\text{max}}]$,
where $\theta^{\text{min}}=0.2$ radian and $\theta^{\text{max}}=0.4$
radian denote the minimum and maximum beamwidths of each SC, respectively.
The transmit power level set of each SC is $\{21,\:23,\:25\}$ dBm
and the SC antenna gain is $5$ dBi. The number of transmit antennas
$N_{b}$ and receive antennas $N_{k}$ at the SC and UE are set to
$64$ and $4$, respectively. The blockage is modeled as a distance-dependent
probability state where the channel is either line-of-sight (LOS)
or non-LOS for urban environments at $28$ GHz and the system bandwidth
is $1$ GHz \cite{2014mmWave_Blockage}. Numerical results are obtained via Monte-Carlo simulations over $50$ different random topologies. The risk-sensitive parameter is set to $\mu_{b}=-2$. For the learning algorithm, the trade-off factor $\kappa_{b}$ is set to $5$, while the learning rates $\zeta_{b}(t)$ and $\iota_{b}(t)$ are set to
$\frac{1}{\left(t+1\right)^{0.55}}$ and $\frac{1}{\left(t+1\right)^{0.6}}$,
respectively \cite{2013learning}. Furthermore, we compare our proposed \textbf{RSL} scheme with the following baselines:
\begin{itemize}
	\item \textbf{\textit{Classical Learning}} (\textbf{CSL}) refers to the
	RL framework in which the utility function only considers the
	mean value of mmWave links \cite{2013learning}.
	\item \textbf{\textit{Baseline 1}}\textbf{ }(\textbf{BL1}) refers to \cite{2017hybrid} optimizing
	the beamwidth with maximum transmit power.
\end{itemize}
In Fig. \ref{CCDF}, we plot the complementary cumulative distribution
function (tail distribution - CCDF) of user throughput (UT) at $28$
GHz when the number of SCs is $24$ per $\text{km}^{2}$. The CCDF
curves reflect the reliable probability (in both linear and logarithmic
scales), defined as the probability that the UT is higher than a target
rate $r_{0}$ Gbps, i.e, $\text{Pr\ensuremath{\left(\text{UT\ensuremath{\geq}r}_{0}\right)}}$. We also study the impact of imperfect CSI with $\tau_k = 0.3$ and feedback with noise from UEs. We observe that the performance of our proposed RSL framework is reduced under these impacts. We next compare our proposed RSL method with other baselines with perfect CSI and user feedback.  It is observed that the \textbf{RSL} scheme achieves better reliability, $\text{Pr\ensuremath{\left(\text{UT\ensuremath{\geq}10 Gbps}\right)}}$, of more than $85\%$, whereas the baselines \textbf{CSL} and \textbf{BL1}
obtain less than $75\%$ and $65\%$, respectively. However, at very
low rate (less than $2$ Gbps) or very high rate ($10.65-11$ Gbps)
captured by the cross-point, the \textbf{RSL} obtains a lower probability as compared to the baselines. In other words, our proposed solution provides a UT which is more concentrated around its median in order to provide uniformly great service for all users. For instance, the UT distribution of our proposed algorithm has a small variance of $0.4846$, while the \textbf{CSL} has a higher variance of $2.6893$.

\begin{figure}[t]
    \centering
   \includegraphics[width=1\columnwidth]{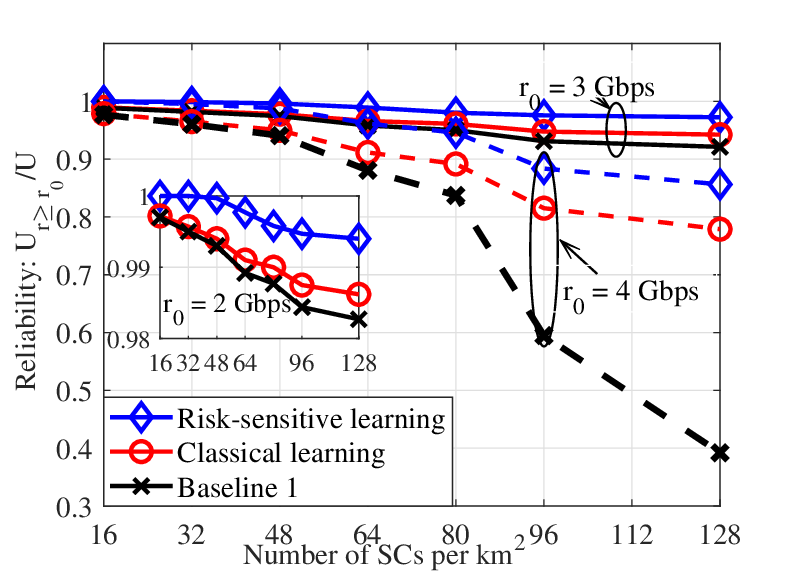}
	\caption{Reliability versus network density.}
    \vspace{-1em}
	\label{relPro}
\end{figure}

\subsection{Impact of network density}
Fig. \ref{relPro} reports the impact of network density on the reliability,
which is defined as the fraction of UEs who achieve a given target
rate $r_{0}$, i.e., $\frac{K_{r>r_{0}}}{K}$. Here, the number of
SCs is varying from $16$ to $128$ per $\text{km}^{2}$. For given
target rates of $2$, $3$, and $4$ Gbps, our proposed algorithm
guarantees higher \textit{reliability} as compared to the baselines.
Moreover, the higher the target rate, the bigger the performance gap
between our proposed algorithm and the baselines. A linear increase
in network density decreases reliability, for example, when the density
increases from $16$ to $96$, the fraction of users that achieve
$4$ Gbps of the \textbf{RSL}, \textbf{CSL}, and \textbf{BL1} are
reduced by $11.61\%$, $16.72\%,$ and $39.11\%$, respectively. This
highlights a key tradeoff between reliability and network density.

\begin{figure}[t]
	\centering
		\includegraphics[width=1\columnwidth]{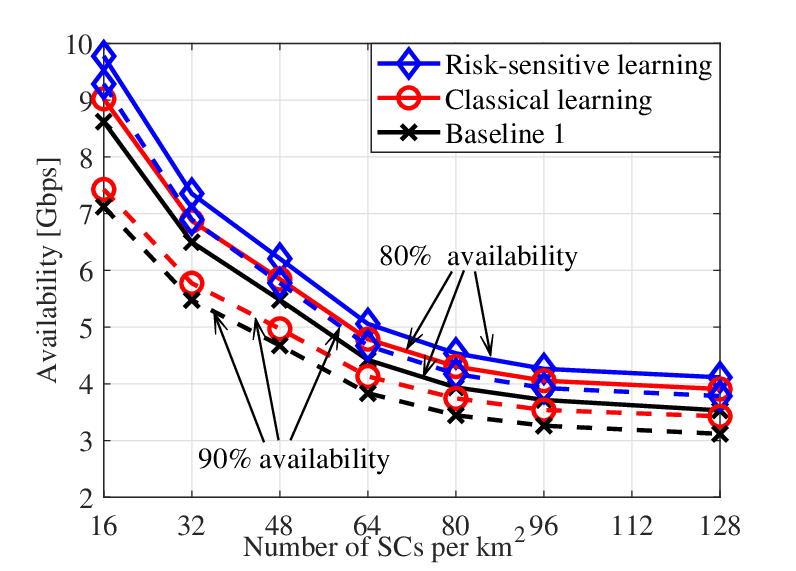}
		\caption{Availability versus network density.}
		\vspace{-1em}
		\label{densityimpact}
	\end{figure}

In Fig. \ref{densityimpact} we show the impact of network
density on the availability, which defines how much rate is obtained
for a target probability. We plot the $80\%$ and $90\%$ probabilities
in which the system achieves a rate of at least $r$ Gbps. For a given
target probability of $90\%$, our proposed algorithm guarantees more
than $9$ Gbps of UT, whereas the baselines guarantee less than $7.5$
Gbps of UT for $B=16$, while if we lower the target probability to
$80\%$, the achievable rate is increased by $5\%$. This gives rise
to a tradeoff between the reliability and the data rate. In addition,
for a given probability, the achievable rate $r$ is reduced with
the increase in network density. For instance, when the network density
increases from $16$ to $80$, the achievable rate is reduced by $50\%$.
This highlights the tradeoff between availability and network density.

\begin{figure}[t]
		\centering
		\includegraphics[width=1\columnwidth]{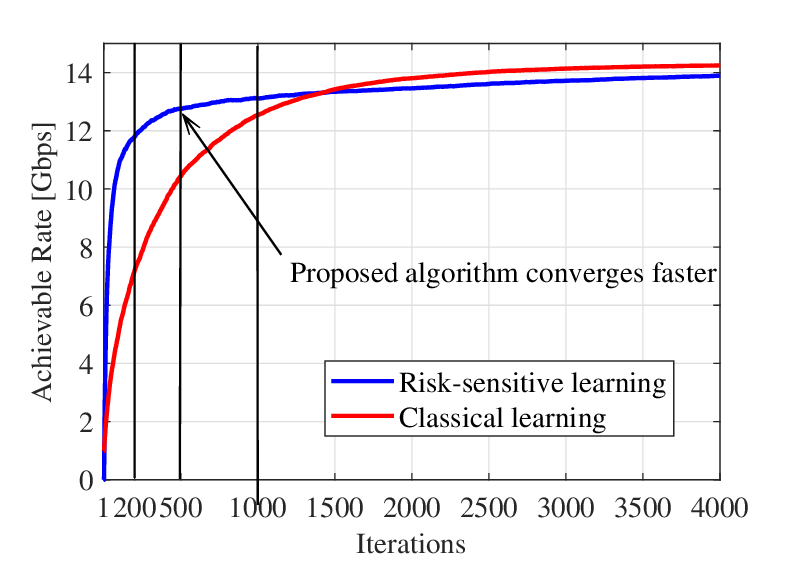}
		\caption{Convergence of the proposed RSL and  classical RL.}
		\vspace{-1em}
		\label{Learning_duration1}
\end{figure}

We numerically observe that $T=4000$ is long enough for agents to learn and enjoy the optimal solution. We assume that the channel condition is changed after every $T=4000$. Our proposed algorithm converges faster than the classical learning baseline as shown in Fig. \ref{Learning_duration1}. By harnessing the notion of risk-averse, the agents try to find the best strategy subject to the variations of the mmWave rates. Basically, the classical RL approach is based upon the exploitation and exploration paradigm, in
which the agents find all possible actions to optimize the expected utility over a given time period. In the
risk-averse case, the agents also try to find the best strategy by taking into account the variations of the
mmWave transmission rates. Hence, the RSL agents do not try to exploit the strategies with either very high
gain or very low gain. While the classical RL exploits all strategies that leads to a longer learning duration.
As can be seen in Fig. \ref{Learning_duration1} the classical RL needs a longer learning duration to find the optimal solution.
In contrast, the RSL obtains a near-optimal solution with a shorter learning duration, while reducing the
variances of the achievable rates.
    \vspace{-1em}
\section{Conclusions}
In this letter, the problem of providing multi-gigabit wireless access with reliable communication was studied by optimizing the transmit beam and considering the link sensitivity in $5$G mmWave networks. A distributed risk-sensitive RL based  approach was proposed taking into account both mean and variance values of the mmWave links. Numerical results show that the proposed approach provides
better services for all users. For instance, the proposed approach
achieves a $\text{Pr\ensuremath{\left(\text{UT\ensuremath{\geq}\ 10Gbps}\right)}}$
is higher than $85\%$, whereas the baselines obtain less than
$75\%$ and $65\%$ with  $24$ small cells.

The proposed reinforcement learning algorithms allow a distributed manner for individual network elements to independently operate. However, the proposed reinforcement learning algorithm works only in static and sparse networks. In a high mobility environment,  a fast convergent solution is required. Together with the problem of beam selection and power allocation, the beam tracking and alignment become more challenging in high mobility mmWave networks. Dynamic networks with high mobility demanding high reliability and low latency require optimal solutions in a reasonable time. In this regard, deep reinforcement learning is a promising solution to obtain a faster convergence speed  and handle a large number of state-action pairs.

Moreover, to solve a problem of a large population or actions space, leveraging tools from mean-field theory or machine learning (i.e.,  actor/critic approaches) can ease the curse of dimensions. 
\bibliographystyle{IEEEtran}
\bibliography{mmWave}
\end{document}